\documentclass[usenatbib]{mn2e}

\usepackage[dvips]{graphicx}
\usepackage{amssymb}
\usepackage{txfonts}
\usepackage{colordvi}

\newcommand{\g}{$\gamma$}
\newcommand{\li}{{\rm Li}_2}

\title[Compton scattering of blackbody photons]{Compton scattering of blackbody photons by relativistic electrons}

\author[A. A. Zdziarski and P. Pjanka]
{Andrzej A. Zdziarski$^1$ and Patryk Pjanka$^2$\\
$^1$Centrum Astronomiczne im.\ M. Kopernika, Bartycka 18, PL-00-716 Warszawa, Poland\\
$^2$Obserwatorium Astronomiczne Uniwersytetu Warszawskiego, Al. Ujazdowskie 4, 00-478 Warszawa, Poland\\
}

\date{Accepted 2013 September 18.  Received 2013 September 18; in original form 2013 July 25 }

\pagerange{\pageref{firstpage}--\pageref{lastpage}}
\pubyear{2013}

\begin{document}

\maketitle

\label{firstpage}

\begin{abstract}
We present simple and accurate analytical formulae for the rates of Compton scattering by relativistic electrons integrated over the energy distribution of blackbody seed photons. Both anisotropic scattering, in which blackbody photons arriving from one direction are scattered by an anisotropic electron distribution into another direction, and scattering of isotropic seed photons are considered. Compton scattering by relativistic electrons off blackbody photons from either stars or CMB takes place, in particular, in microquasars, colliding-wind binaries, supernova remnants, interstellar medium, and the vicinity of the Sun.
\end{abstract}
\begin{keywords}
acceleration of particles -- gamma-rays: general -- gamma-rays: stars -- radiation mechanisms: non-thermal -- relativistic processes -- scattering
\end{keywords}

\section{Introduction}
\label{intro}

Compton scattering by relativistic electrons is a very common astrophysical process. Electrons are accelerated to relativistic energies in shocks or reconnection in a variety of cosmic sources, e.g., jets, colliding winds in binaries, supernova remnants, or coronae of accreting sources. Relativistic electrons are also a component of cosmic rays. The seed photons are often blackbody, usually from either the cosmic microwave background (CMB) or stars. The blackbody field is anisotropic if it comes from a star or isotropic, e.g., in the CMB. Some sites where this process takes place are listed below.

Black-hole X-ray binaries in some of their states have jets, which contain relativistic electrons. In the case of binaries with massive donors, e.g., Cyg X-1 or Cyg X-3, the electrons significantly Compton up-scatter stellar radiation (see, e.g, \citealt{jackson72,gak02,brp06,fermi,dch10b,zls12,z12,zps13,mzc13}). The blackbody radiation field coming from the surface of the donor star arriving at the jet is strongly anisotropic. The probability of scattering is highest for head-on collisions, and relativistic electrons radiate in a beam along their direction. This leads to a pronounced anisotropy of the Compton-scattered radiation, e.g., \citet{jackson72}, \citet*{dch08,dch10a,dch10b}, \citet{dubus13}, with the observed Compton flux peaking around the superior conjunction. Assuming the star to be a point source, calculations of the phase-averaged jet emission require 4-dimensional integration, over the electron and photon distributions, the jet length and the orbital phase. Taking into account the finite size of the star adds two more dimensions to the integration. The distribution of the relativistic electrons is, in general, not a power law, in particular when energy losses and a high-energy cut-off are considered (e.g., \citealt*{zps13}). Then Compton rates averaged over power law electrons cannot be used. Such calculations are thus very computationally intensive, and integrating analytically the Compton rate over the blackbody distribution significantly facilitates them. 

The same Compton anisotropy takes place in binary systems with pulsar wind colliding with wind of a massive star, so called \g-ray binaries (see \citealt{dubus13} for a review). In those systems, interaction of the two winds leads to acceleration of relativistic electrons, which then radiate, in particular Compton up-scatter blackbody photons from the massive star. That radiation will then be orbitally modulated, e.g., in PSR B1259--63, LS I +61$\degr$ 303 and LS 5039 \citep*{kbs99,dch08,cerutti10}.

Related systems are so-called colliding wind binaries, comprising of two massive stars with interacting winds. Electrons are accelerated in the interaction region, and Compton up-scatter the stellar radiation. Such emission appears to be observed from $\eta$ Car \citep{bp11}.

In the case of pulsar winds, stellar photons may also be Compton up-scattered by cold electrons in the wind, which is ultra-relativistic and can have the bulk Lorentz factor as high as $\gamma\ga 10^5$. This process takes place in \g-ray binaries \citep*{cdh08}, but it appears to be seen also in systems with pulsars with low-mass companions, e.g., PSR B1957+20, the black widow pulsar \citep{wu12}. In these cases, both the electron distribution and the blackbody radiation from the star are anisotropic. A related process is Compton up-scattering by relativistic bulk motion of electrons in a jet, considered by \citet{bs87}. 

Another astrophysical case of anisotropic Compton scattering is that of cosmic-ray electrons and the interstellar starlight, which contributes to the Galactic diffuse emission. That starlight field is dominated by stars in the Galactic plane, and thus it is anisotropic \citep{ms00}. Furthermore, cosmic-ray electrons in the solar system Compton up-scatter (anisotropic) solar radiation, which gives rise to emission of \g-rays \citep*{mpd06,os07,os08}. (A software package for Compton scattering by relativistic electrons off stellar emission is given by \citealt{os13}.)

Cosmic-ray electrons in the Galaxy also up-scatter the isotropic CMB, contributing to X-ray and \g-ray backgrounds, e.g., \citet{fm66}, \citet{strong75}. This process also takes place in other galaxies, see, e.g., \citet{cf04}, \citet{smail12}.

Then, Compton up-scattering by relativistic electrons of both CMB and starlight takes place in supernova remnants and pulsar wind nebulae, e.g., \citet{lazendic04}, \citet*{pms06}, \citet{hess11}. Usually the starlight field is approximated as isotropic.

Here we study both anisotropic and isotropic scattering. We integrate the corresponding rates of Compton scattering by mono-energetic electrons over the blackbody distribution, obtaining simple and highly accurate formulae in close form. In the isotropic case, we compare our results with the approximation of \citet{petruk09}, as well as with calculation in the electron rest frame, see Appendix \ref{rest}. 

Our results give Compton photon redistribution functions for highly relativistic electrons at a single energy, i.e., the distribution of the energies and directions of the scattered photon. At low energies, the electrons have usually a Maxwellian distribution, for which the corresponding redistribution function has recently been presented by \citet*{spw12}.

\section{Integration of the Compton cross section over a blackbody spectrum}
\label{integration}

\subsection{Anisotropic seed photons}
\label{aniso}

The rate of Compton scattering (in both Thomson and Klein-Nishina regimes) of a photon beam from direction $\mathbf{\Omega}$ into direction $\mathbf{\Omega_1}$ by a cloud of relativistic electrons with Lorentz factors $\gamma\gg 1$ is given by equation (20) of \citet{aa81}, and it can be written as,
\begin{eqnarray}
\lefteqn{
{{\rm d} \dot n\over {\rm d}\epsilon_1 {\rm d}\epsilon {\rm d}\gamma {\rm d}\Omega_1 {\rm d}\Omega}= {3\sigma_{\rm T}c \over 4\epsilon\gamma^2} {{\rm d}N(\gamma)\over {\rm d}\Omega_1} {{\rm d}n_0(\epsilon)\over {\rm d}\Omega} \left[1+\epsilon_1 \epsilon_{\rm m} y-{2\epsilon_{\rm m}\over \epsilon}+{2\epsilon_{\rm m}^2\over \epsilon^2}\right],\nonumber}\\
\lefteqn{\quad
\quad y\equiv 1-\mathbf{\Omega}\cdot \mathbf{\Omega_1},\quad \epsilon_{\rm m}\equiv {\epsilon_1\over 2 y\gamma (\gamma-\epsilon_1)}, \label{rate}}
\end{eqnarray}
where ($\epsilon$, $\mathbf{\Omega}$) and ($\epsilon_1$, $\mathbf{\Omega_1}$) are the energy in units of $m_{\rm e}c^2$ and the direction of the incoming and scattered photon, respectively, $m_{\rm e}$ is the electron mass, ${\rm d}N(\gamma)/{\rm d}\Omega_1$ is the electron number per unit $\gamma$ and per unit solid angle, and ${\rm d}n_0/{\rm d}\Omega$ is the density of photons coming from direction of $\mathbf{\Omega}$. Then, the range of kinematically allowed values of $\epsilon$, $\epsilon_1$ and $\gamma$ can be expressed in three equivalent forms,
\begin{equation}
\left[\epsilon\geq \epsilon_{\rm m},\, \epsilon_1<\gamma\right],\,\, \epsilon_1\leq {2 y\epsilon\gamma^2\over 1+2 y\epsilon\gamma},\,\, \gamma\geq 
{\epsilon_1\over 2} \left(1+\sqrt{ 1+{2\over \epsilon\epsilon_1 y}}\right), 
\label{range}
\end{equation}
as well as $\epsilon\ll \epsilon_1$ is required. The first of the above conditions is used for integration over $\epsilon$. Note that equation (\ref{rate}) assumes the directions of the electron and scattered photons to be the same. This is because electrons with $\gamma\gg 1$ emit in a narrow beam along its direction of motion, and thus, in the adopted approximation, only electrons moving along the scattered photon direction contribute to the spectrum. Consequently, the same equation applies for either anisotropic or isotropic ${\rm d}N(\gamma)/{\rm d}\Omega_1$ (although \citealt{aa81} considered it for isotropic electrons only). Indeed, equation (\ref{rate}) agrees, in the limits of $\gamma\gg 1$ and $\epsilon\ll \epsilon_1$, with equation (20) of \citet*{fks97}, which gives a formula for emission by an electron beam scattering off a photon beam valid at any $\gamma$ and $\epsilon$. For an isotropic electron distribution, ${\rm d}N(\gamma)/{\rm d}\Omega_1= N(\gamma)/(4\upi)$. We note that our equation (\ref{rate}), equation (20) of \citet{fks97}, as well as equation (6) of \citet{os08} are given for unit ${\rm d}N(\gamma)/{\rm d}\Omega_1$ whereas equation (20) of \citet{aa81} is normalized to $N(\gamma)$. 

\begin{figure}
\centerline{\includegraphics[width=7.5cm]{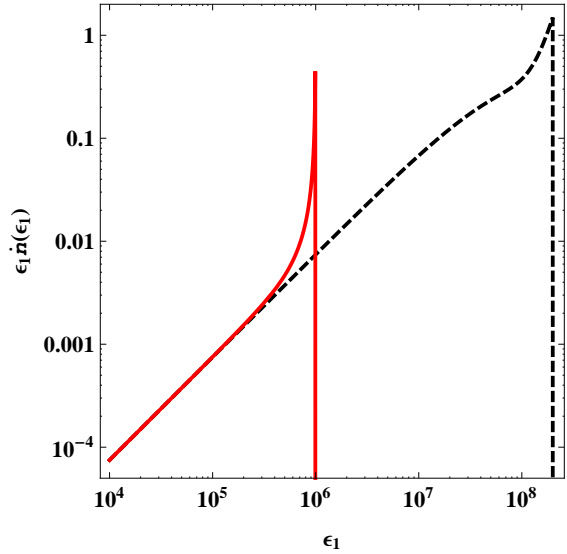}}
\caption{The scattering rate on seed photons at a single energy, equation (\ref{rate}), plotted per unit $\ln \epsilon_1$ (red solid curve) compared to the Thomson-limit rate (black dashed curve) for $\epsilon=10^{-4}$, $\gamma=10^6$, $y=1$, and in the Thomson units. At low $\epsilon_1$, the rate per unit $\epsilon_1$ is constant.
} \label{ratekn_th}
\end{figure}

In the Thomson limit, in which the photon energy in the electron rest frame, $\epsilon_*\equiv y\epsilon\gamma$, is $\ll 1$, the rate (\ref{rate}) is given by the corresponding expression without the term $\epsilon_1 \epsilon_{\rm m}y$ and with $\epsilon_{\rm m}=\epsilon_1/(2 y \gamma^2)$. On the other hand, the low-energy, $\epsilon_1\ll 2y \epsilon\gamma^2/(1+2y \epsilon\gamma)$, limit of the scattering rate is different, and given by equation (\ref{rate}) with the factor in brackets replaced by 1, which is valid for {\it any\/} $\epsilon_*$. The distinction between these two limits is illustrated in Fig.\ \ref{ratekn_th}.

An interesting feature of the rate (\ref{rate}) in the Klein-Nishina limit, $\epsilon_*\gg 1$, is it showing a sharp peak at the maximum allowed $\epsilon_1$, above which it is cut off. This is illustrated in Fig.\ \ref{ratekn_th}, comparing that rate to its Thomson limit. The total, integrated over $\epsilon_1$, rate is lower, but the rate per unit $\epsilon_1$ is much higher than that in the Thomson limit just before the cutoff. The presence of the peak is due to a compression of the allowed range of $\epsilon_1$ close to $\gamma$ in the Klein-Nishina limit, since the condition of $\epsilon_1<\gamma$ has to be satisfied for any scattering. Mathematically, the peak appears due to the presence of the $(\gamma-\epsilon_1)$ factor in the denominator of the definition of $\epsilon_{\rm m}$.

The total scattering rate, i.e., that of equation (\ref{rate}) integrated over $\epsilon_1$, is directly related to the total Klein-Nishina cross section (e.g., \citealt{rl79}), $\sigma_{\rm KN}$,
\begin{equation}
{{\rm d} \dot n\over {\rm d}\epsilon {\rm d}\gamma {\rm d}\Omega_1 {\rm d}\Omega}= y\sigma_{\rm KN}(\epsilon_*)c {{\rm d}N(\gamma)\over {\rm d}\Omega_1} {{\rm d}n_0(\epsilon)\over {\rm d}\Omega}.
\label{total_rate}
\end{equation}
In the Thomson limit, the rate averaged over the the cosine of the scattering angle is given by $\sigma_{\rm T}c$. 

\begin{figure}
\centerline{\includegraphics[width=7.5cm]{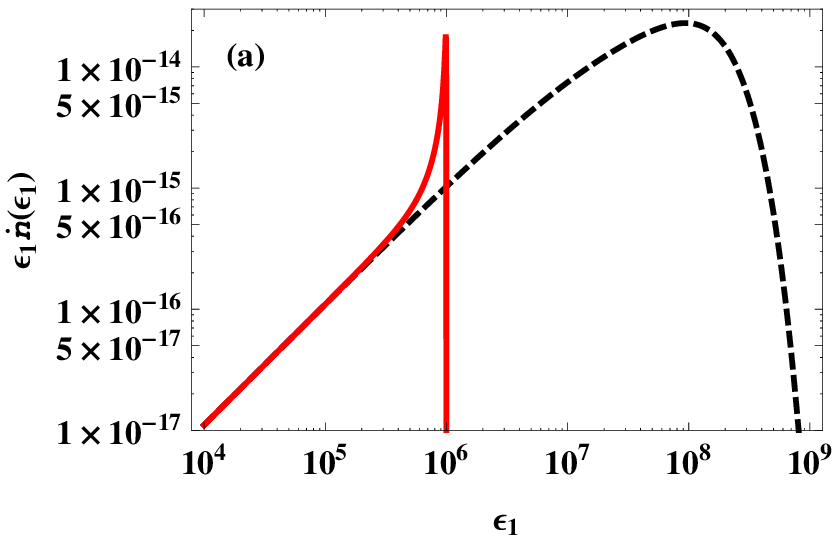}}
\centerline{\includegraphics[width=7.5cm]{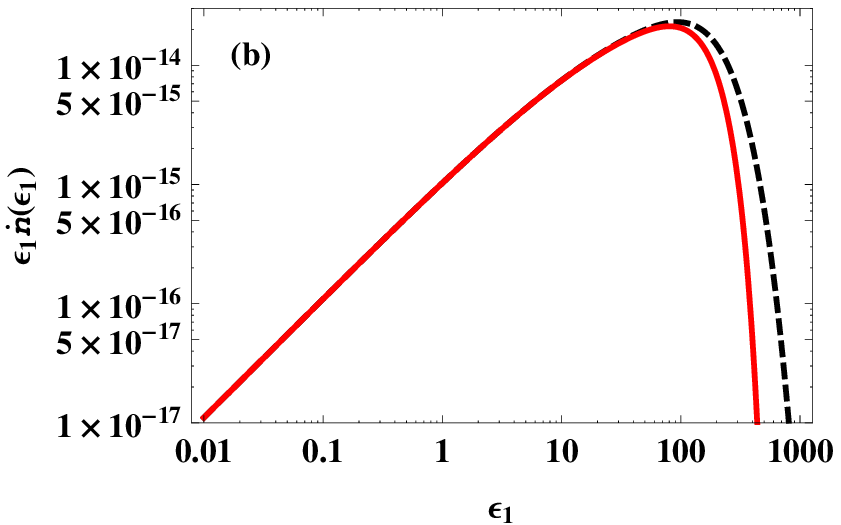}}
\caption{The blackbody-integrated scattering rate of equation (\ref{ratebb}) per unit $\ln \epsilon_1$ (red solid curves) compared to the Thomson-limit rate (black dashed curves) for $\theta=3\times 10^{-5}$, $y=1$ and (a) $\gamma=10^6$ and (b) $\gamma=10^3$. The units are Thomson and $A=1$. The maximum values of $\epsilon_1$ corresponding to the average blackbody energy, $\epsilon\simeq 2.7\theta$, given by equation (\ref{range}), are $\simeq 10^6$ and $\simeq 140$ for (a) and (b), respectively.
} \label{ratebbkn_th}
\end{figure}

A direction-dependent diluted blackbody photon density can be written as,
\begin{equation}
{{\rm d}n_0(\epsilon)\over {\rm d}\Omega}= {A(\mathbf{\Omega}) \epsilon^2 \over \exp(\epsilon /\theta)-1},
\label{bb}
\end{equation}
where $\theta=k T/m_{\rm e} c^2$, $T$ is the temperature, $A(\mathbf{\Omega})$ is direction-dependent normalization, and $T$ and $\epsilon$ are given in the frame of the scattering gas. Note that the relation to the specific intensity, $I$, is $I(\epsilon, \Omega)=c \epsilon\, m_{\rm e} c^2{{\rm d}n_0(\epsilon)/{\rm d}\Omega}$. For undiluted blackbody, $A=2 (m_{\rm e}c/h)^3$, where $h$ is the Planck constant. The case of photons arriving from a single direction, $\mathbf{\Omega_0}$, corresponds to the appearance of $\delta(\mathbf{\Omega}- \mathbf{\Omega_0})$ in $A(\mathbf{\Omega})$. Then, the scattering rate can integrated over $\Omega$. In particular, for photons arriving from a spherical star with radius $R_*$ at a distance from its centre of $R\gg R_*$, in which case the arriving photons can be approximated by a mono-directional beam, we have,
\begin{equation}
A(\mathbf{\Omega})=2\upi{\cal D}_*\left(m_{\rm e}c\over h\right)^3 \left(R_*\over R\right)^2 \delta(\mathbf{\Omega}- \mathbf{\Omega_0}),
\label{star}
\end{equation}
where ${\cal D}_*$ is the Doppler factor of stellar photons seen in the electron gas frame. This formula follows from the fact that the flux, $F$, from a uniformly emitting sphere is given by $F=\upi I (R_*/R)^2$ \citep{rl79}, and the photon density for a mono-directional beam is $F/(c \epsilon\, m_{\rm e}c^2)$. 

The scattering rate can be integrated over the photon distribution,
\begin{equation}
{{\rm d} \dot n\over {\rm d}\epsilon_1 {\rm d}\gamma {\rm d}\Omega_1 {\rm d}\Omega}= \int_{\epsilon_{\rm m}}^{\infty} {{\rm d} \dot n(\epsilon)\over {\rm d}\epsilon_1 {\rm d}\epsilon {\rm d}\gamma {\rm d}\Omega_1{\rm d}\Omega} {\rm d}\epsilon,
\label{int_rate}
\end{equation}
Here, we have set the upper limits of integration to infinity for the sake of simplicity, in spite of the condition of $\epsilon\ll \epsilon_1$ assumed in equation (\ref{rate}). This has a negligible effect on the results as long as $\theta\ll \epsilon_1$. 

Comparing equations (\ref{rate}) and (\ref{bb}), we see that the rate (\ref{int_rate}) involves three integrals with different powers of $\epsilon$. We can perform two of them analytically,
\begin{eqnarray}
\lefteqn{ 
\int\limits_x^\infty \!\! {{\rm d}x'\,x'\over \exp x'-1}\equiv f_1(x)={\upi^2\over 6}+{x^2 \over 2}+{\rm Li}_2\left(1-\exp x\right)=\nonumber}\\
\lefteqn{\qquad
{\rm Li}_2\left[\exp(- x)\right]-x \ln\left[1- \exp\left(-x\right)\right],\label{int1}}\\
\lefteqn{
\int\limits_x^\infty  {{\rm d}x'\over \exp x'-1}\equiv f_0(x)= -\ln\left[1- \exp\left(-x\right)\right],\label{int2}}
\end{eqnarray}
where ${\rm Li}_2(x)$ is the dilogarithm ($=-\int_0^x {\rm d}x' \ln(1-x')/x'$; note that \citealt{petruk09} uses a different definition). 

The third integral has no closed form. We thus use the following series expansions,
\begin{equation}
{1\over \exp x-1}=x^{-1}\!-{1\over 2}+\sum_{k=1}^{N=\infty} {x^{2 k-1}B_{2 k}\over (2 k)!},\quad x<2\upi,
\label{series1}
\end{equation}
where $B_i$ is the Bernoulli number  \citep{gr80}, and
\begin{equation}
{1\over \exp x-1}=\sum_{k=1}^{N=\infty} \exp\left(-k x\right),\quad x>0.
\label{series2}
\end{equation}
Since $B_{2 k}/ (2 k)!$ decreases very fast, in particular, it is equal to $1/12$, $-1/720$ and $1/30240$ for $k=1$, 2, 3, respectively, and the second series is a sum of exponentials of negative numbers, both series are rapidly converging. We can integrate the series (\ref{series1}) times $x^{-1}$ term by term for $x<2\upi$, which yields
\begin{equation}
\int  {{\rm d}x x^{-1}\over \exp x-1}\simeq g_{\rm l}(x,N)= -x^{-1}-{\ln x\over 2}+\sum_{k=1}^{N}{ x^{2 k-1} B_{2 k}\over (2k-1)(2 k)!,
\label{int3l}}
\end{equation}
which is exact for $N\rightarrow \infty$. We then calculate the definite integral of $g_{\rm l}(q, N)-g_{\rm l}(x, N)+g_0$ with $q=2.257$, $N=3$, where
\begin{equation}
g_0=\int_{q}^\infty {{\rm d}x\,x^{-1}\over \exp x-1}\simeq 0.0366377.
\label{int3c}
\end{equation}
We can then integrate the series (\ref{series2}) times $x^{-1}$ term by term,
\begin{equation}
\int_{x}^\infty  {{\rm d}x' x'^{-1}\over \exp x'-1}\simeq g_{\rm h}(x,N)=\sum_{k=1}^{N} {\rm E}_1\!\left(k x\right),\quad x>0,
\label{int3h}
\end{equation}
which is exact for $N\rightarrow \infty$, and where ${\rm E}_1(x)\,\left( = \int_1^\infty {\rm d} t\, {\rm e}^{-x t}/t\right)$ is the exponential integral. Our approximation to the definite integral, equation (\ref{int3h}), is then,
\begin{equation}
f_{-1}(x)\equiv \cases{g_{\rm l}(q , 3)-g_{\rm l}(x, 3)+g_0, &$x\leq q$;\cr
g_{\rm h}(x, 3), &$x> q$,}\quad q=2.257.
\label{f-1}
\end{equation}
The maximum fractional error of this expression for any value of $x$ is $<3.6\times 10^{-4}$, which error is reached around $x=q$. We chose the value of $q$ to minimize the maximum error of our approximation.

The resulting rate integrated over blackbody is,
\begin{eqnarray}
\lefteqn{{{\rm d} \dot n\over {\rm d}\epsilon_1 {\rm d}\gamma{\rm d}\Omega_1{\rm d}\Omega}= {3 \sigma_{\rm T}c A \over 4 \gamma^2}{{\rm d}N(\gamma)\over {\rm d}\Omega_1} \left[\left(1+ \epsilon_1\epsilon_{\rm m} y \right)\theta^2 f_1\left(\epsilon_{\rm m}\over \theta\right) 
+\right.\nonumber}\\
\lefteqn{
\quad \left.  +2\epsilon_{\rm m}^2 f_{-1}\left(\epsilon_{\rm m}\over \theta\right)-2\epsilon_{\rm m}\theta f_0\left(\epsilon_{\rm m}\over \theta\right)
\right]. \label{ratebb}}
\end{eqnarray}
In the Thomson regime of $2.7\theta\gamma\ll 1$ (where $2.7\theta$ is the average blackbody photon energy), the rate is as above but without the $\epsilon_1\epsilon_{\rm m} y$ term and with $\epsilon_{\rm m}=\epsilon_1/(2 y \gamma^2)$. The $f$-function argument, $\epsilon_{\rm m}/\theta$, can still be $\ga 1$. On the other hand, in the low-energy limit of $\epsilon_1\ll 2y \gamma^2 \theta/(1+2y \gamma \theta)$, $\dot n$ is constant with $\epsilon_1$,
\begin{equation}
{{\rm d} \dot n\over {\rm d}\epsilon_1 {\rm d}\gamma{\rm d}\Omega_1{\rm d}\Omega}= {\sigma_{\rm T}c A \over \gamma^2}{{\rm d}N(\gamma)\over {\rm d}\Omega_1} {\upi^2 \theta^2\over 8},
\label{ratebbt}
\end{equation}
for any value of $\theta\gamma$, from the Thomson to extreme Klein-Nishina regimes. This constant rate is seen in Figs.\ \ref{ratebbkn_th}(a--b), which show examples of $\dot n$ in the Klein-Nishina and Thomson regimes. It also shows the corresponding Thomson-limit dependencies. In the Klein-Nishina regime, Fig.\ \ref{ratebbkn_th}(a), we see a sharp peak around the maximum allowed $\epsilon_1$, similar to that in Fig.\ \ref{ratekn_th}.

\subsection{Isotropic seed photons}
\label{iso}

In the case of isotropic seed photons, $n_0(\epsilon)= 4\upi\, {{\rm d}n_0(\epsilon)/ {\rm d}\Omega}$, the scattering rate has been derived by \citet{jones68}, see also equation (2.48) of \citet{bg70} or equation (22) of \citet{aa81}. The scattering rate into a given direction, $\mathbf{\Omega_1}$, is 
\begin{eqnarray}
\lefteqn{{{\rm d} \dot n\over {\rm d}\epsilon_1 {\rm d}\epsilon {\rm d}\gamma {\rm d}\Omega_1}= {3\sigma_{\rm T}c n_0(\epsilon) \over 4\epsilon\gamma^2}\! {{\rm d}\!N(\gamma)\over {\rm d}\Omega_1} \left[1+ 2\epsilon_1\epsilon_{\rm n} +{\epsilon_{\rm n}(1-2\epsilon_1\epsilon_{\rm n})\over \epsilon} 
-{2\epsilon_{\rm n}^2\over \epsilon^2}\right.\nonumber}\\
\lefteqn{
\quad \left. +{2\epsilon_{\rm n}\over \epsilon}\ln{\epsilon_{\rm n}\over \epsilon}
\right],\quad \epsilon_{\rm n}\equiv {\epsilon_1\over 4\gamma (\gamma-\epsilon_1)}. \label{rateiso}}
\end{eqnarray}
As required, an integral of equation (\ref{rate}) over $\Omega$ (i.e., over $y$ and times $2\upi$) equals the above formula. Note that since relativistic electrons emit mostly along their direction of motion, the above formula applies for anisotropic electrons. If the electrons are isotropic, we can integrate over $\Omega_1$ (i.e., multiply by $4\upi$) and substitute ${\rm d}N(\gamma)/{\rm d}\Omega_1=N(\gamma)/(4\upi)$ in equations (\ref{rateiso}), (\ref{rateisobb}--\ref{rateisobbp}). (Thus, only removal of $\Omega_1$ on the left-hand side and replacement of ${\rm d}N(\gamma)/{\rm d}\Omega_1$ by $N(\gamma)$ is required.) The ranges of kinematically allowed values of $\epsilon$, $\epsilon_1$ and $\gamma$ are given by $\epsilon\ll \epsilon_1$ and
\begin{equation}
\left[\epsilon\geq \epsilon_{\rm n},\,\, \epsilon_1<\gamma\right],\,\, \epsilon_1\leq {4\epsilon\gamma^2\over 1+4\epsilon\gamma},\,\,\gamma\geq 
{\epsilon_1\over 2} \left(1+\sqrt{ 1+{1\over \epsilon\epsilon_1}}\right). 
\label{rangeiso}
\end{equation}
The rate in the Thomson limit, $\epsilon\gamma\ll 1$, corresponds to the rate (\ref{rateiso}) without the terms of $2\epsilon_1 \epsilon_{\rm n}$ and with $\epsilon_{\rm n}=\epsilon_1/(4\gamma^2)$. The low-energy rate is constant, analogously to the case of anisotropic scattering, with the term in brackets set to 1. In the Klein-Nishina limit, the profile of $\dot n$ shows a sharp peak very similar to that shown in Fig.\ \ref{ratekn_th}.

The total scattering rate, i.e., integrated over $\epsilon_1$, is given by \citep{akv85,protheroe86,z88}
\begin{eqnarray}
\lefteqn{
{{\rm d} \dot n\over {\rm d}\epsilon {\rm d}\gamma {\rm d}\Omega_1}=
{3\sigma_{\rm T}c n_0(\epsilon)\over 2 s^2}{{\rm d}N(\gamma)\over {\rm d}\Omega_1}\left[\left(s+9+{8\over s}\right)\ln(1+s) +\right. \nonumber}\\
\lefteqn{\qquad
-{16+18s+s^2\over 2(1+s)} +4\li(-s)\bigg]\label{int_iso}}\\
\lefteqn{
\qquad \simeq \sigma_{\rm T}c n_0(\epsilon){{\rm d}N(\gamma)\over {\rm d}\Omega_1}\left[1-{2 s\over 3}+{13 s^2\over 20} +O(s^3)\right],}
\end{eqnarray}
where $s=4\epsilon\gamma$ is twice the maximum electron rest-frame seed photon energy. This rate can also be obtained by averaging equation (\ref{total_rate}) over $y$. The corresponding electron energy loss rate, $\dot \gamma$, is given by \citet{jones68} (see also \citealt{z88}).

The angle-integrated density of isotropic photons is,
\begin{equation}
n_0(\epsilon)= 4\upi {{\rm d}n_0(\epsilon)\over {\rm d}\Omega}={A \epsilon^2 \over \exp(\epsilon /\theta)-1},
\label{bbiso}
\end{equation}
where $A=8\upi (m_{\rm e} c/h)^3$ for undiluted blackbody. The integrated scattering rate is,
\begin{equation}
{{\rm d} \dot n\over {\rm d}\epsilon_1 {\rm d}\gamma {\rm d}\Omega_1}= \int_{\epsilon_{\rm n}}^{\infty} {{\rm d} \dot n(\epsilon)\over {\rm d}\epsilon_1 {\rm d}\epsilon {\rm d}\gamma{\rm d}\Omega_1} {\rm d}\epsilon.
\label{int_rate_iso}
\end{equation}

\begin{figure}
\centerline{\includegraphics[width=7.5cm]{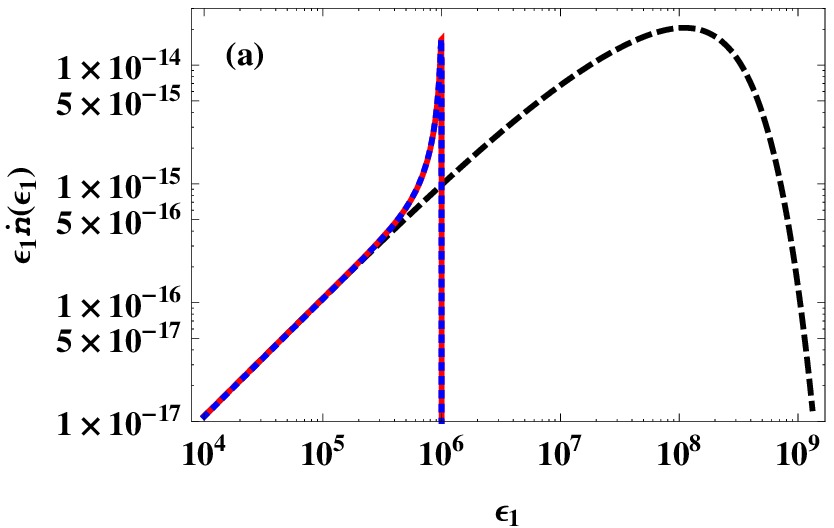}}
\centerline{\includegraphics[width=7.5cm]{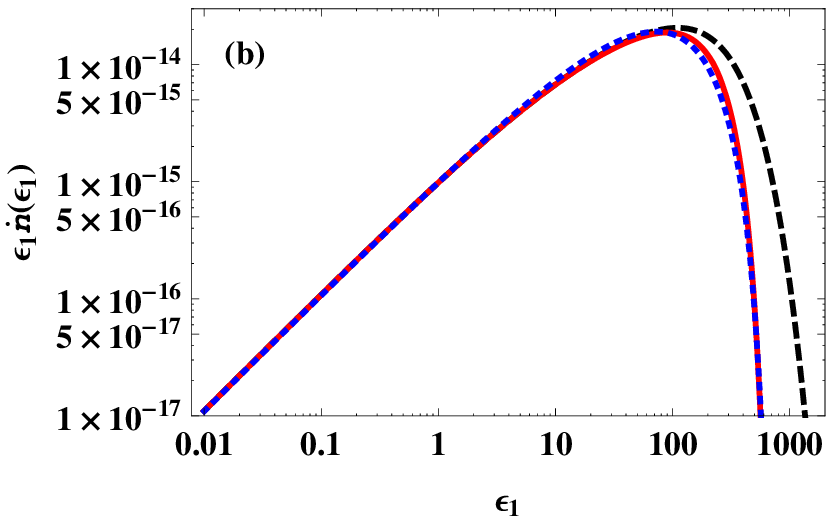}}
\centerline{\includegraphics[width=7cm]{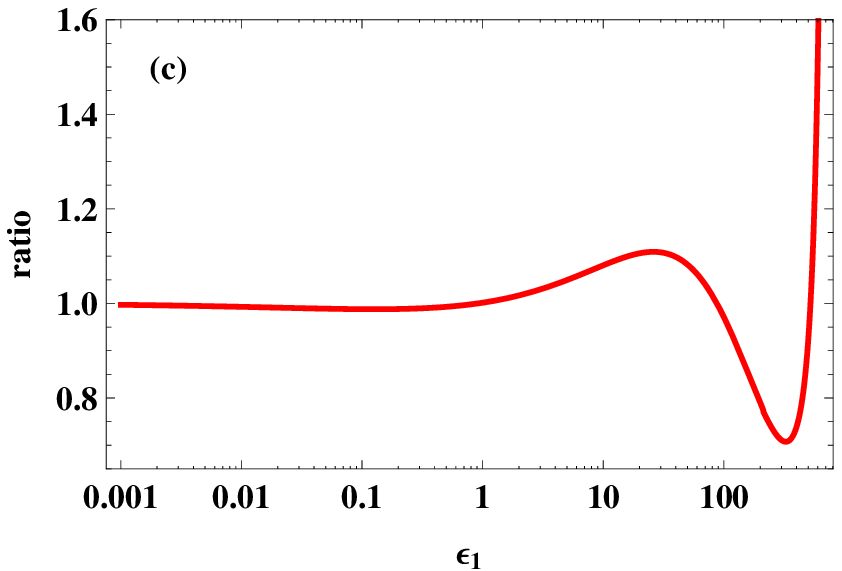}}
\caption{The isotropic blackbody-integrated scattering rate of equation (\ref{rateisobb}) per unit $\ln \epsilon_1$ (red solid curves) compared to the approximation of \citet{petruk09} (blue dotted curves) and the Thomson-limit rate (black dashed curves) for $\theta=3\times 10^{-5}$, and (a) $\gamma=10^6$ and (b) $\gamma=10^3$. The units are Thomson and $A=1$. (c) The ratio of the approximation of \citet{petruk09} to the actual rate for the case shown in the panel (b). 
} \label{ratebbiso63}
\end{figure}

For this rate with blackbody photons, we have the same powers of $\epsilon$ as in the anisotropic case, and, in addition, an integral involving $\ln\epsilon$. We can integrate the series (\ref{series1}) times $\ln x$ term by term, which yields,
\begin{eqnarray}
\lefteqn{
\int {{\rm d}x\,\ln x\over \exp x-1}\simeq h_{\rm l}(x,N),\quad x<2\upi,}\\
\lefteqn{
h_{\rm l}(x,N)={x(1-\ln x)+\ln^2 x\over 2}+\sum_{k=1}^{N} x^{2 k}\!{(2k\ln x-1)B_{2 k}\over 4 k^2(2 k)!},
\label{intlog1}}
\end{eqnarray}
which is exact for $N\rightarrow \infty$. We calculate the definite integral of $h_{\rm l}(q, N)-h_{\rm l}(x, N)+h_0$ with $q$ and $N$ as above, where
\begin{equation}
h_0=\int_{q}^\infty {{\rm d}x\,\ln x\over \exp x-1}\simeq 0.125505.
\label{intlogc}
\end{equation}
We then integrate the series (\ref{series2}) times $\ln x$ term by term,
\begin{eqnarray}
\lefteqn{
\int_{x}^\infty {{\rm d}x' \ln x'\over \exp x'-1}\simeq h_{\rm h}(x, N),\quad x>0,\label{intln}}\\
\lefteqn{
h_{\rm h}(x,N)=
\sum_{k=1}^{N} {{\rm E}_1\! \left({k x}\right)+\exp\left({-k x}\right)\ln x \over k},
\label{intlog2}}
\end{eqnarray}
which is exact for $N\rightarrow \infty$. Our approximation to the definite integral, equation (\ref{intln}), is then,
\begin{equation}
f_{\ln}(x)\equiv \cases{h_{\rm l}(q, 3)-h_{\rm l}(x, 3)+h_0, &$x\leq q$;\cr
h_{\rm h}(x, 3), &$x> q$,}\quad q=2.257.
\label{fln}
\end{equation}
We note that the definite integral, see equation (\ref{intln}), is equal to 0 at $x_0\simeq 0.438849$. Then, the fractional error of any approximation of it, in particular that of equation (\ref{fln}), will be large close to $x_0$. However, since the value of the integral is then $\simeq 0$, the contribution of this error is negligible. Apart from that, the maximum fractional error of equation (\ref{fln}) is $<2.4\times 10^{-4}$ (reached around $x=q$). The accuracy can be arbitrarily improved by increasing $N$.

The resulting blackbody-integrated rate is,
\begin{eqnarray}
\lefteqn{{{\rm d} \dot n\over {\rm d}\epsilon_1 {\rm d}\gamma {\rm d}\Omega_1}= {3 \sigma_{\rm T}c A \over 4\gamma^2} {{\rm d}N(\gamma)\over {\rm d}\Omega_1} \left[ \left(1-2\epsilon_1\epsilon_{\rm n}+ 2\ln{\epsilon_{\rm n}\over \theta}\right)\epsilon_{\rm n}\theta f_0\left(\epsilon_{\rm n}\over \theta\right) 
+\right.\nonumber}\\
\lefteqn{
\quad \left. +\left(1+ 2\epsilon_1\epsilon_{\rm n}\right)\theta^2 f_1\left(\epsilon_{\rm n}\over \theta\right) -2\epsilon_{\rm n}^2 f_{-1}\left(\epsilon_{\rm n}\over \theta\right)
-2 \epsilon_{\rm n}\theta f_{\rm ln}\left(\epsilon_{\rm n}\over \theta\right)
\right]. \label{rateisobb}}
\end{eqnarray}
The Thomson regime, $2.7\theta\gamma\ll 1$, has the rate as above but without the terms of $2\epsilon_1\epsilon_{\rm n}$ and with $\epsilon_{\rm n}=\epsilon_1/ (4\gamma^2)$. On the other hand, the low-energy, $\epsilon_1\ll 4\gamma^2 \theta/(1+4\gamma \theta)$, limit has a constant rate in any regime, where the factor in brackets above is replaced by $\theta^2 \upi^2/6$, similar to that of equation (\ref{ratebbt}), see Fig.\ \ref{ratebbiso63}(a--b). We note that the integration over the blackbody spectrum can also be done in the electron rest frame, see Appendix \ref{rest}, which gives identical results to the numerical integration (\ref{int_rate_iso}).

The fractional error of our equations (\ref{ratebb}) and (\ref{rateisobb}) is $<10^{-5}$ over most of the parameter space. The highest error of $\simeq 6\times 10^{-4}$ for equation (\ref{ratebb}) is reached around $\epsilon_{\rm m}/\theta\simeq q$ in the Thomson regime, and the error is much lower in the Klein-Nishina regime. Equation (\ref{rateisobb}) has the highest error around $\epsilon_{\rm n}/\theta\simeq q$, reaching $\simeq 3\times 10^{-3}$ in the Thomson regime. 

\citet{petruk09} obtained an approximate formula for the Compton rate integrated over isotropic blackbody photons, given by their equation (32), which, in our notation, is
\begin{eqnarray}
\lefteqn{{{\rm d} \dot n\over {\rm d}\epsilon_1 {\rm d}\gamma{\rm d}\Omega_1}= {\sigma_{\rm T}c A \theta^2\upi^2 \over 8\gamma^2}{{\rm d}N(\gamma)\over {\rm d}\Omega_1} \exp\left(-{2\epsilon_{\rm n}\over 3\theta}\right)\times
\nonumber}\\
\lefteqn{
\quad \left\{ \exp\left[-{5\over 4}\left(\epsilon_{\rm n}\over \theta\right)^{1/2}\right]+ 2\epsilon_1 \epsilon_{\rm n}
\exp\left[-{5\over 7}\left(\epsilon_{\rm n}\over \theta\right)^{7/ 10}\right]\right\}.
 \label{rateisobbp}}
\end{eqnarray}
Fig.\ \ref{ratebbiso63}(a--b) shows the rate of equation (\ref{rateisobb}) compared to its Thomson limit and to the approximation of \citet{petruk09}. We find the formula of \citet{petruk09} to be generally quite accurate. Fig.\ \ref{ratebbiso63}(c) shows its accuracy in a Thomson-limit case. Equation (\ref{rateisobbp}) is fully accurate in the low-energy limit, and then it has the fractional error of $\la 30$ per cent at all energies except the high-energy tail, where, however, the scattering rate itself is negligibly small, see the middle panel. 

\section{Conclusions}
\label{discussion}

We have obtained simple and accurate formulae for the rates of Compton scattering by relativistic electrons of a given energy off blackbody photons. The blackbody-integrated rate for scattering from one direction into another is given by equation (\ref{ratebb}). The corresponding rate for scattering off isotropic blackbody seed photons is given by equation (\ref{rateisobb}). These formulae apply to electrons with either isotropic or anisotropic angular distribution. We note that the isotropic calculations apply also to the case of a mono-directional photon beam irradiating isotropic electrons provided the spectrum of the scattered photons is integrated over all directions.

Our results involve two special functions, the dilogarithm and the exponential integral. Those functions can be easily calculated in any standard numerical package, e.g., \citet{numrec}, \citet{wolfram96}. Obviously, the required integrations can also be carried out numerically. However, formulae are always preferable to numerical results. In particular, using our formulae in multidimensional integration of jet emission \citep{zps13} instead of integrating the scattering rate over blackbody has resulted in a very substantial increase of computational efficiency, in some cases reducing the computing time by a factor of $\sim 100$. 

\section*{ACKNOWLEDGMENTS}

We thank M. Sikora, A. Strong, W. Bednarek, and J. Poutanen for valuable discussions and suggestions. We also thank the referee for valuable suggestions. This research has been supported in part by the Polish NCN grants N N203 581240 and 2012/04/M/ST9/00780.

\appendix
\section{Scattering in the electron rest frame}
\label{rest}

We note that \citet{strong75} considered isotropic scattering by relativistic electrons on blackbody photons using an approach different from that adopted in Section \ref{iso}. He calculated the spectrum of isotropic blackbody photons in the electron rest frame (see also see also \citealt{fs98}), assuming that all the photons arrive in the frame direction, which is approximately satisfied for the electron Lorentz factor of $\gamma\gg 1$. This yields \citep{strong75},
\begin{equation}
n_0(\epsilon_*)=\left(m_{\rm e}c\over h \right)^3 {4\upi \epsilon_*\theta\over \gamma^2} \ln {1\over 1-\exp(-\epsilon_* /2\gamma\theta)},
\label{bbrest}
\end{equation}
where $\epsilon_*$ is the dimensionless photon energy in the electron rest frame, see Section \ref{aniso}.

Since $n_0(\epsilon)$ is an invariant (e.g., \citealt{bg70}), the above distribution can be used for numerical integration of the rate of Compton scattering over $\epsilon_*$, see equation (4.7) in \citet{strong75}. This is fully equivalent to integrating the angle-averaged rate (\ref{rateiso}) with blackbody photons over $\epsilon$, see Section \ref{iso}. An advantage of using the photon distribution in the particle rest frame is that it can be readily used for other reactions by highly relativistic particles on blackbody photons, e.g., e$^\pm$ pair production by cosmic-ray protons.

\label{lastpage}


\begin{thebibliography}{}

\bibitem[\protect\citeauthoryear{Abdo et al.}{2009}]{fermi} 
Abdo, A. A., et al., 2009, Sci, 326, 1512

\bibitem[\protect\citeauthoryear{Aharonian \& Atoyan}{1981}]{aa81} 
Aharonian F.~A., Atoyan A.~M., 1981, Ap\&SS, 79, 321 

\bibitem[\protect\citeauthoryear{Aharonian et al.}{1985}]{akv85} 
Aharonian F. A., Kirillov-Ugryumov V. G., Vardanian, V. V., 1985, Ap\&SS, 115, 201

\bibitem[\protect\citeauthoryear{Bednarek \& Pabich}{2011}]{bp11} 
Bednarek W., Pabich J., 2011, A\&A, 530, A49 

\bibitem[\protect\citeauthoryear{Begelman \& Sikora}{1987}]{bs87} 
Begelman M.~C., Sikora M., 1987, ApJ, 322, 650 

\bibitem[\protect\citeauthoryear{Blumenthal \& Gould}{1970}]{bg70} 
Blumenthal G.~R., Gould R.~J., 1970, RvMP, 42, 237 

\bibitem[\protect\citeauthoryear{Bosch-Ramon, Romero \& Paredes}{Bosch-Ramon et al.}{2006}]{brp06} 
Bosch-Ramon V., Romero G.~E., Paredes J.~M., 2006, A\&A, 447, 263 

\bibitem[\protect\citeauthoryear{Celotti \& Fabian}{2004}]{cf04} 
Celotti A., Fabian A.~C., 2004, MNRAS, 353, 523 

\bibitem[\protect\citeauthoryear{Cerutti, Dubus \& Henri}{Cerutti et al.}{2008}]{cdh08} 
Cerutti B., Dubus G., Henri G., 2008, A\&A, 488, 37 

\bibitem[\protect\citeauthoryear{Cerutti et al.}{2010}]{cerutti10} 
Cerutti B., Malzac J., Dubus G., Henri G., 2010, A\&A, 519, A81 

\bibitem[\protect\citeauthoryear{Dubus}{2013}]{dubus13} 
Dubus G., 2013, A\&ARv, 21, 64

\bibitem[\protect\citeauthoryear{Dubus, Cerutti \& Henri}{Dubus et al.}{2008}]{dch08} 
Dubus G., Cerutti B., Henri G., 2008, A\&A, 477, 691 

\bibitem[\protect\citeauthoryear{Dubus, Cerutti \& Henri}{Dubus et al.}{2010a}]{dch10a} 
Dubus G., Cerutti B., Henri G., 2010a, A\&A, 516, A18 

\bibitem[\protect\citeauthoryear{Dubus, Cerutti \& Henri}{Dubus et al.}{2010b}]{dch10b} 
Dubus G., Cerutti B., Henri G., 2010b, MNRAS, 404, L55

\bibitem[\protect\citeauthoryear{Fargion \& Salis}{1998}]{fs98} 
Fargion D., Salis A., 1998, PhyU, 41, 823 

\bibitem[\protect\citeauthoryear{Fargion, Konoplich \& Salis}{Fargion et al.}{1997}]{fks97} 
Fargion D., Konoplich R.~V., Salis A., 1997, ZPhyC, 74, 571 

\bibitem[\protect\citeauthoryear{Felten \& Morrison}{1966}]{fm66} 
Felten J.~E., Morrison P., 1966, ApJ, 146, 686 

\bibitem[\protect\citeauthoryear{Georganopoulos, Aharonian \& Kirk}{Georganopoulos et al.}{2002}]{gak02}
{Georganopoulos} M., {Aharonian} F.~A., {Kirk} J.~G., 2002, A\&A, 388, L25

\bibitem[\protect\citeauthoryear{Gradshteyn \& Ryzhik}{1980}]{gr80} 
Gradshteyn I.~S., Ryzhik I.~M., 1980, Table of integrals, series and products. New York: Academic Press  

\bibitem[\protect\citeauthoryear{H.E.S.S.~Collaboration}{2011}]{hess11} H.E.S.S.~Collaboration, 2011, A\&A, 533, A103 

\bibitem[\protect\citeauthoryear{Jackson}{1972}]{jackson72} 
Jackson J.~C., 1972, Nat.\ Phys.\ Sci., 236, 39 

\bibitem[\protect\citeauthoryear{Jones}{1968}]{jones68} 
Jones F.~C., 1968, PhRv, 167, 1159 

\bibitem[\protect\citeauthoryear{Kirk, Ball \& Skjaeraasen}{Kirk et al.}{1999}]{kbs99} 
Kirk J.~G., Ball L., Skjaeraasen O., 1999, APh, 10, 31 

\bibitem[\protect\citeauthoryear{Lazendic et al.}{2004}]{lazendic04} 
Lazendic J.~S., Slane P.~O., Gaensler B.~M., Reynolds S.~P., Plucinsky P.~P., Hughes J.~P., 2004, ApJ, 602, 271 

\bibitem[\protect\citeauthoryear{Malyshev, Zdziarski \& Chernyakova}{Malyshev et al.}{2013}]{mzc13} 
Malyshev D., Zdziarski A.~A., Chernyakova M., 2013, MNRAS, 434, 2380

\bibitem[\protect\citeauthoryear{Moskalenko \& Strong}{2000}]{ms00} 
Moskalenko I.~V., Strong A.~W., 2000, ApJ, 528, 357

\bibitem[\protect\citeauthoryear{Moskalenko, Porter \& Digel}{Moskalenko et al.}{2006}]{mpd06} 
Moskalenko I.~V., Porter T.~A., Digel S.~W., 2006, ApJ, 652, L65 

\bibitem[\protect\citeauthoryear{Orlando \& Strong}{2007}]{os07} 
Orlando E., Strong A.~W., 2007, Ap\&SS, 309, 359 

\bibitem[\protect\citeauthoryear{Orlando \& Strong}{2008}]{os08} 
Orlando E., Strong A.~W., 2008, A\&A, 480, 847 

\bibitem[\protect\citeauthoryear{Orlando \& Strong}{2013}]{os13} 
Orlando E., Strong A. W., 2013, Nuc.\ Phys.\ Suppl., 239, 266 

\bibitem[\protect\citeauthoryear{Petruk}{2009}]{petruk09} 
Petruk O., 2009, A\&A, 499, 643 

\bibitem[\protect\citeauthoryear{Porter, Moskalenko \& Strong}{Porter et al.}{2006}]{pms06} 
Porter T.~A., Moskalenko I.~V., Strong A.~W., 2006, ApJ, 648, L29 

\bibitem[\protect\citeauthoryear{Press et al.}{1992}]{numrec} 
Press W.~H., Teukolsky S.~A., Vetterling W.~T., Flannery B.~P., 1992, Numerical recipes in C. The art of scientific computing. Cambridge University Press

\bibitem[\protect\citeauthoryear{Protheroe}{1986}]{protheroe86}
Protheroe R. J., 1986, MNRAS, 221, 769

\bibitem[\protect\citeauthoryear{Rybicki \& Lightman}{1979}]{rl79}
Rybicki G.~B., Lightman A.~L., 1979, Radiative Processes in Astrophysics. 
New York: Wiley

\bibitem[\protect\citeauthoryear{Smail et al.}{2012}]{smail12}
Smail I., Blundell K.~M., Lehmer B.~D., Alexander D.~M., 2012, ApJ, 760, 
132 

\bibitem[\protect\citeauthoryear{Strong}{1975}]{strong75} 
Strong, A. W., 1975, PhD thesis, University of Durham, http://www.mpe.mpg.de/$\sim$aws/publications/Thesis.pdf

\bibitem[\protect\citeauthoryear{Suleimanov, Poutanen \& Werner}{Suleimanov et al.}{2012}]{spw12} 
Suleimanov V., Poutanen J., Werner K., 2012, A\&A, 545, A120 

\bibitem[\protect\citeauthoryear{Wolfram}{1996}]{wolfram96} 
Wolfram S., 1996, The Mathematica book. Cambridge University Press

\bibitem[\protect\citeauthoryear{Wu et al.}{2012}]{wu12} 
Wu E.~M.~H., Takata J., Cheng K.~S., Huang R.~H.~H., Hui C.~Y., Kong A.~K.~H., Tam P.~H.~T., Wu J.~H.~K., 2012, ApJ, 761, 181 

\bibitem[\protect\citeauthoryear{Zdziarski}{1988}]{z88} 
Zdziarski, A. A., 1988, ApJ, 335, 786

\bibitem[\protect\citeauthoryear{Zdziarski, Lubi{\'n}ski \& Sikora}{Zdziarski et al.}{2012a}]{zls12} 
Zdziarski A.~A., Lubi{\'n}ski P., Sikora M., 2012a, MNRAS, 423, 663

\bibitem[\protect\citeauthoryear{Zdziarski et al.}{2012b}]{z12} 
Zdziarski A.~A., Sikora M., Dubus G., Yuan F., Cerutti B., Ogorza{\l}ek A., 2012b, MNRAS, 421, 2956

\bibitem[\protect\citeauthoryear{Zdziarski, Pjanka \& Sikora}{Zdziarski et al.}{2013}]{zps13} 
Zdziarski A.~A., Pjanka P., Sikora M., 2013, MNRAS, submitted, arXiv:1307.1309

\end{thebibliography}
\end{document}